\documentclass[page-classic]{epl2}
\usepackage{blindtext}
\usepackage{subcaption}
\usepackage{amsmath}
\usepackage{color} 
\usepackage[table]{xcolor}
\usepackage{graphicx}
\usepackage{parskip}

\title{From empirical brain networks towards modeling music perception -- a perspective}
\shorttitle{Towards modeling music perception}

\author{Jakub Sawicki\inst{1,2,3,*}}
\shortauthor{J. Sawicki}

\institute{                    
  \inst{1} Potsdam Institute for Climate Impact Research\\
  \inst{2} Berlin University of the Arts \\
  \inst{3} University of Applied Sciences Northwestern Switzerland \\
  * zergon@gmx.net
}


\abstract{This perspective article investigates how auditory stimuli influence neural network dynamics using the FitzHugh-Nagumo (FHN) model and empirical brain connectivity data. Results show that synchronization is sensitive to both the frequency and amplitude of auditory input, with synchronization enhanced when input frequencies align with the system's intrinsic frequencies. Increased stimulus amplitude broadens the synchronization range governed by a delicate interplay involving the network's topology, the spatial location of the input, and the frequency characteristics of the cortical input signals. This perspective article also reveals that brain activity alternates between synchronized and desynchronized states, reflecting critical dynamics and phase transitions in neural networks. Notably, gamma-band synchronization is crucial for processing music, with coherence peaking in this frequency range. The findings emphasize the role of structural connectivity and network topology in modulating synchronization, providing insights into how music perception engages brain networks. This perspective article offers a computational framework for understanding neural mechanisms in music perception, with potential implications for cognitive neuroscience and music psychology.
}

\begin{document}

\maketitle

\section{Synchronization phenomena in neural dynamics}

Synchronization is a fundamental property of neural dynamics, supporting a wide range of physiological, pathological, and cognitive processes. It is prominently observed during slow-wave sleep and in transitions between wakefulness and sleep \cite{STE93b,RAT00,SCH08o,MOR12}, while partial synchronization has been linked to phenomena such as the first-night effect and unihemispheric sleep, where hemispheric asymmetries in neural synchrony emerge \cite{TAM16,RAT00,RAT16,MAS16,RAM19}. Excessive synchronization underlies certain pathological states such as epileptic seizures, where localized hyper-synchrony disrupts normal brain function \cite{GER20}. Conversely, synchronization facilitates essential cognitive functions, including the development and perception of syntax \cite{KOE13,LAR15a,BAD20}.

The human brain is a hierarchically organized system, with the cortex responsible for higher-order functions such as perception, decision-making, and motor control, and subcortical structures, including the basal ganglia, thalamus, and limbic system, governing motor regulation, emotion, and memory. Coherent interaction between these regions, mediated by neuronal synchronization, supports adaptive behavior and efficient cognitive processing. Functional MRI studies show that increasing motor task complexity results in the recruitment of broader brain networks, with simple rhythmic synchronization tasks engaging contralateral sensorimotor and cerebellar regions, and complex tasks involving additional structures such as the basal ganglia and prefrontal cortex \cite{MAY02a}.

Oscillatory synchronization, particularly in the beta and gamma frequency ranges (20–80 Hz), underpins key neural processes, including attentional modulation and motor coordination \cite{SIN93,UHL09}. These oscillations arise from cortico-cortical and cortico-subcortical interactions and are shaped by conduction delays and developmental plasticity \cite{LOE92}. In atypical populations, such as individuals with autism, altered synchronization patterns have been linked to compensatory engagement of alternative neural pathways, reflecting the brain's capacity to reorganize in response to connectivity disruptions \cite{JUS04a}.

In music perception, synchronization manifests as widespread coherence across cortical areas in response to auditory stimuli \cite{BAD20,SAW22}. Electroencephalography (EEG) studies have shown that music induces event-related potentials (ERPs) with synchronized neural activity in multiple regions, particularly in the beta and gamma bands \cite{HAR14,HAR20a,BAD13,SAW21a}. These rhythms support attentional engagement and the segmentation of musical structure, with synchronization theory offering a robust framework for interpreting these processes in the context of hierarchical auditory perception \cite{JOR94,BAD13,SAW18a,HOU20,SHA20,SAW23b}.

To model these dynamics, we employ the FitzHugh-Nagumo (FHN) oscillator system, a minimal model of neuronal excitability that captures the essential features of spike generation and recovery \cite{BAS18}. Simulations incorporate empirical structural connectivity derived from diffusion-weighted MRI, enabling exploration of how rhythmic auditory inputs propagate through large-scale brain networks. A key phenomenon investigated is tonal fusion, the perception of a unified pitch from a complex overtone spectrum \cite{SCH18k}, modeled using a single-frequency input with fixed amplitude. This abstraction isolates the fundamental entrainment mechanisms operating in oscillator networks.

Empirical coupling matrices based on human brain connectivity \cite{SAN15a,SKO22} define the interactions between nodes in the network, with spatial regions assigned according to the 90-area Automated Anatomical Labeling (AAL) atlas. While this model does not explicitly account for distance-dependent transmission delays, their role in large-scale synchronization is well established \cite{SCH08,PET19}, and future extensions may include them. Relay synchronization mechanisms, where distant regions achieve synchrony via a mediating relay node, further illustrate how network topology supports long-range coordination \cite{LEY18,BER12,NIC13,GAM13,ZHA17,ZHA17a}. Studies on three-layer multiplex networks have shown that such structures can support chimera states and partial synchronization through relay dynamics \cite{SAW18c,SAW18,SAW20,WIN19,DRA20,SAW21}.

Beyond auditory perception, synchronization is integral to sensory-motor integration and perception-action coupling. Magnetoencephalography (MEG) studies show that weak sensory stimuli evoke enhanced phase synchronization in delta/theta (3–7 Hz) and gamma (40–60 Hz) bands, linking sensory, motor, and attentional systems \cite{HIR18}. Synchronization strength correlates with behavioral response times, demonstrating its functional relevance in cognitive tasks. These findings are supported by computational work showing that temporal alignment of neural oscillations enhances mutual information and stabilizes perception by reducing neural noise \cite{BAH24}.

The principle of ``synchrony through synaptic plasticity'' posits that precisely timed inputs -- regardless of their origin -- are essential for neural computation, enabling functions such as signal compression and sparsening. This principle is exemplified in cochlear implant research, where sound-coding strategies aimed at restoring synchrony improve pitch discrimination and speech perception, emphasizing the broader role of temporal structure in auditory cognition.

This perspective article advances the understanding of partial synchronization in neural networks by incorporating empirically derived structural connectivity. Synchronization and coherence are central to brain function, supporting processes across perceptual, cognitive, and pathological domains. In the context of auditory perception, increased coherence between brain dynamics and auditory input underscores the pivotal role of synchronization in sensory processing. By focusing on minimal models capable of generating realistic synchronization patterns, this approach facilitates a precise examination of the neural mechanisms underlying auditory perception and offers a tractable framework for translating theoretical principles into experimental and clinical applications \cite{HAR14,HAR20a}.

First, the effects of an external auditory stimulus on a network of FitzHugh-Nagumo oscillators are explored using diffusion-weighted MRI-based structural connectivity from healthy individuals. This perspective article demonstrates that synchronization patterns can be modulated through systematic variation of the frequency and amplitude of the auditory input, revealing flexible control mechanisms that govern neural coherence. The next section extends this analysis by examining how synchronization depends on network topology, the spatial origin of the stimulus, and the spectral properties of cortical input. This section elucidates the complex dependencies between structural features and stimulus characteristics in shaping synchronization dynamics. In the following two sections, the investigation shifts to musical input, revealing that specific musical pieces can significantly enhance coherence between network activity and auditory stimuli. This effect illustrates the brain's capacity for frequency-specific alignment with external rhythmic structures and emphasizes the functional relevance of particular frequency bands in auditory-driven neural synchronization.

Overall, this work integrates phenomenons of complex systems into computational models of perception, highlighting their importance for understanding both fundamental neural mechanisms and the brain's dynamic responsiveness to structured external input. These insights contribute to the development of more comprehensive models of sensory processing and have implications for applications such as brain-computer interfaces and auditory-based therapeutic interventions.

\section{Empirical brain networks}
\label{empirical}

\begin{figure}[tp]
    \begin{center}
\includegraphics[width = .6\linewidth]{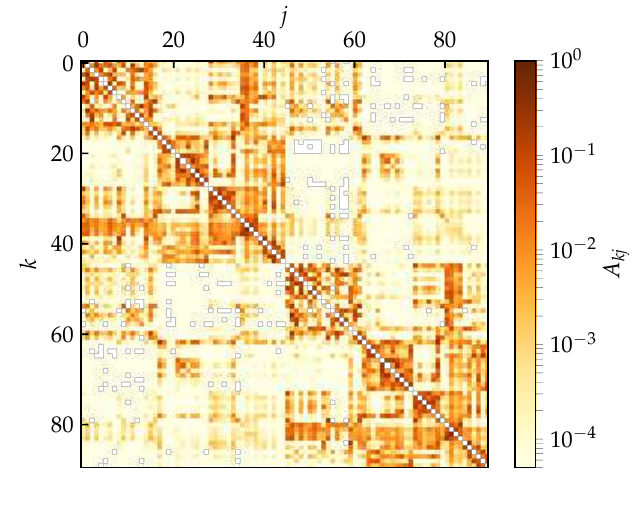}
    \caption{(color online) Model for the brain structure: Weighted adjacency matrix $A_{kj}$ of the averaged empirical structural brain network derived from twenty healthy human subjects by averaging over the coupling between two brain regions $k$ and $j$. The brain regions $k,j$ are taken from the Automated Anatomic Labeling atlas \cite{TZO02}, but re-labeled such that $k=1,...,45$ and $k=46,...,90$ correspond to the left and right hemisphere, respectively.
    After \cite{GER20,SKO22}.}
    \label{fig.1}
    \end{center}
\end{figure}

We consider an empirical structural brain network shown in Fig.\,\ref{fig.1} where every region of interest is modeled by a single FitzHugh-Nagumo (FHN) oscillator. The auditory cortex is the part of the temporal lobe that processes auditory information in humans. It is a part of the auditory system, performing basic and higher functions in hearing and is located bilaterally, roughly at the upper sides of the temporal lobes, i.e., corresponding to the AAL indexing $k = 41,86$ (temporal sup L/R) in Tab.\,\ref{cha5:SAW24:tab:AAL}. The auditory cortex takes part in the spectrotemporal analysis of the inputs passed on from the ear.

Each node corresponding to a brain region is modeled by the FitzHugh-Nagumo (FHN) model with external stimulus, a paradigmatic model for neural spiking \cite{FIT61,NAG62,BAS18}. Note that while the FitzHugh-Nagumo model is a simplified model of a single neuron, it is also often used as a generic model for excitable media on a coarse-grained level \cite{CHE05e,CHE07a}. The FHN dynamics of the network with external stimulus at a specific pair of cortical regions reads:

\begin{subequations}
\begin{align}
\epsilon \dot{u}_k = &u_k - \frac{u_k^3}{3} - v_k \nonumber \\
                    &+ \sigma \sum_{j \in N_\text{H}} A_{kj} \left[ B_{uu}(u_j - u_k) + B_{uv}(v_j - v_k) \right]  \\
                    &+ \varsigma \sum_{j \notin N_\text{H}} A_{kj} \left[ B_{uu}(u_j - u_k) + B_{uv}(v_j - v_k) \right], \nonumber \\
                    & + I(t) \nonumber\\
\dot{v}_k = &u_k + a \nonumber \\
 & + \sigma \sum_{j \in N_\text{H}} A_{kj} \left[ B_{vu}(u_j - u_k) + B_{vv}(v_j - v_k) \right]  \\
 & + \varsigma \sum_{j \notin N_\text{H} } A_{kj} \left[ B_{vu}(u_j - u_k) + B_{vv}(v_j - v_k) \right], \nonumber
\end{align}
\label{cha5:SAW22:eq.1}
\end{subequations}
with $k \in N_\text{H}$ where $N_\text{H}$ denotes either the set of nodes $k$ belonging to the left ($N_L$) or the right ($N_R$) hemisphere. Parameter $\epsilon = 0.05$ describes the timescale separation between the fast activator variable (neuron membrane potential) $u$ and the slow inhibitor (recovery variable) $v$ \cite{FIT61}. Depending on the threshold parameter $a$, the FHN model may exhibit excitable behavior ($\left| a \right| > 1$) or self-sustained oscillations ($\left| a \right| < 1$). We use the FHN model in the oscillatory regime and thus fix the threshold parameter at $a=0.5$ sufficiently far from the Hopf bifurcation point. 
The coupling within the hemispheres is given by the coupling strength $\sigma$ while the coupling between the hemispheres is given by the inter-hemispheric coupling strength $\varsigma$. The interaction scheme between nodes is characterized by a rotational coupling matrix:

\begin{equation}
\mathbf{B} = 
\begin{pmatrix}
B_{uu} & B_{uv} \\
B_{vu} & B_{vv}
\end{pmatrix}
=
\begin{pmatrix}
\text{cos}\phi & \text{sin}\phi \\
-\text{sin}\phi & \text{cos}\phi
\end{pmatrix},
\end{equation}
with coupling phase $\phi = \frac{\pi}{2} - 0.1$, causing primarily an activator-inhibitor cross-coupling. This particular scheme was shown to be crucial for the occurrence of partial synchronization patterns in ring topologies \cite{OME13} as it reduces the stability of the completely synchronized state. Also in the modeling of epileptic-seizure-related synchronization phenomena \cite{GER20}, where a part of the brain synchronizes, it turned out that such a cross-coupling is important. The subtle interplay of excitatory and inhibitory interaction is typical of the critical state at the edge of different dynamical regimes in which the brain operates \cite{MAS15a,SHI22}, and gives rise to partial synchronization patterns which are not found otherwise.

\section{Influence of sound on empirical brain networks} %
\label{cha5:SAW21a}

This section examines the influence of an external auditory stimulus on a network of FitzHugh-Nagumo oscillators, using empirically derived structural connectivity from diffusion-weighted MRI data of healthy human subjects. This perspective article systematically analyzes how variations in the frequency and amplitude of the sound input modulate synchronization dynamics within the network. Results indicate that synchronization is enhanced when the input frequency resonates with the intrinsic frequencies of individual oscillators or the collective resonance of the network, while amplitude further modulates the spatial extent and degree of synchrony.

The analysis highlights frequency as a primary control parameter, with amplitude acting as a secondary modulator, demonstrating the network's sensitivity to auditory input. A minimal computational model is employed to abstract auditory-driven neural dynamics, leveraging the FitzHugh-Nagumo framework to explore the emergence of partially synchronized states -- conditions posited to support sensory integration and cognitive processes such as attention and memory.

Building on prior research linking neural synchronization to musical perception \cite{SAW21a}, this section suggests that auditory input may serve as a functional probe of brain connectivity, particularly in auditory-related regions. Here, the external stimulus $I(t)=C_k\gamma \cos \omega t$ is modeled by a trigonometric function  with frequency $\omega$ and amplitude $\gamma$ and is applied to the brain areas $k=41,86$ associated with the auditory cortex, i.e. $C_k=1$ if $k=41$ or $86$ and zero otherwise. The coupling between the single regions is given by the coupling strength $\sigma,\varsigma$. As we are looking for partial synchronization patterns we fix $\sigma = \varsigma = 0.6$ similar to numerical studies of synchronization phenomena during unihemispheric sleep \cite{RAM19} and epileptic seizures \cite{GER20} where partial synchronization patterns have been observed.

\begin{figure}
\centering
\includegraphics[width = .9\textwidth]{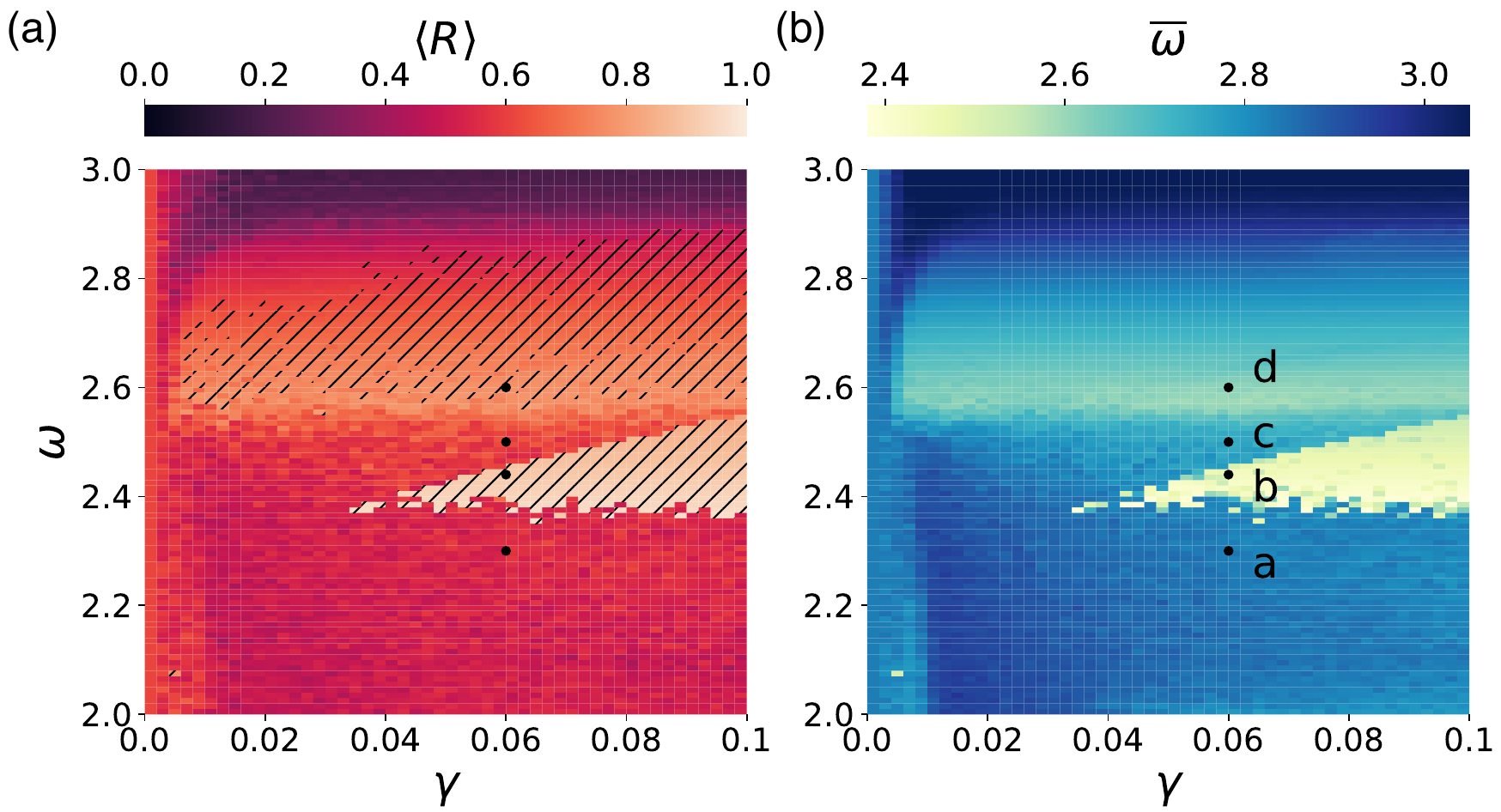}
    \caption{Synchronization tongues in brain network with external stimulus: (a) The temporal mean of the Kuramoto order parameter $\langle R \rangle$ for simulation time $\Delta T=10\,000$ and (b) the spatially averaged mean phase velocity $\overline{\omega}$ in the parameter plane of the frequency $\omega$ of the external stimulus and its amplitude $\gamma$. The light color in panel (a) stands for synchronization and the darker color for desynchronization. In the hatched region the standard deviation of $\langle R \rangle$ is less than $0.1$, which indicates the absence of strong fluctuations of $R$ in time. Other parameters are given by $\sigma=\varsigma=0.6$, $\epsilon = 0.05$, $a=0.5$, and $\phi = \frac{\pi}{2} - 0.1$. Figure taken from \cite{SAW21a}.}
    \label{cha5:SAW21a:fig.2}
\end{figure}

We investigate synchronization scenarios emerging from an external periodic stimulus in the auditory cortices of both hemispheres ($k=41,86$). Figure~\ref{cha5:SAW21a:fig.2} shows synchronization scenarios of an empirical structural brain network in dependence of the frequency $\omega$ and amplitude $\gamma$ of the external stimulus. The light colored regions in Fig.\,\ref{cha5:SAW21a:fig.2}a indicates synchronized dynamics, whereas the darker colors indicate desynchronized dynamics. There is a light colored stripe for $\omega=2.6$ which indicates a Kuramoto order parameter $\langle R \rangle \approx 0.8$ and a light colored tongue starting at $\omega=2.4, \gamma=0.04$. The hatched region in Fig.\,\ref{cha5:SAW21a:fig.2}a stands for a low standard deviation $<0.1$ of the temporal mean of the Kuramoto order parameter $\langle R \rangle$. It indicates the absence of strong fluctuations of $R(t)$ and therefore a constant high level of synchrony in time. Figure~\ref{cha5:SAW21a:fig.2}b shows the drop of the spatially averaged mean phase velocity $\overline{\omega}$ in case of coherent dynamics in the synchronization regions of Fig.\,\ref{cha5:SAW21a:fig.2}a. In the upper region, $\overline{\omega}$ takes over the value of the frequency $\omega$ of the external stimulus, whereas in the synchronization tongue $\overline{\omega}$ keeps its value of $\overline{\omega}=2.4$. 

It turns out that by taking the frequency $\omega$ of the external stimulus as a control parameter, one can change the level of synchrony of the system. Fixing the external driving amplitude at \(\gamma = 0.06\), the temporal evolution of the Kuramoto order parameter \(R\) and the mean phase velocities of the nodes are analyzed for varying external frequencies \(\omega\). At \(\omega = 2.3\), the system exhibits large temporal fluctuations in \(R\), resembling the unstimulated regime. Notably, only the auditory cortex entrains to the external frequency, while other regions maintain higher intrinsic frequencies around \(\omega_k \approx 2.8\).

A slight increase to \(\omega = 2.4\) induces a transition to global synchrony, characterized by a high order parameter (\(R \approx 0.95\)) and uniform mean phase velocities matching the collective frequency \(\Omega\). Further increase to \(\omega = 2.5\) disrupts synchrony, placing the system in a transition region between two synchronization domains. At \(\omega = 2.6\), close to the uncoupled natural frequency, partial synchronization re-emerges (\(R \approx 0.8\)). However, a hemispheric asymmetry in mean phase velocities appears: the right hemisphere is coherent, while the left exhibits desynchronization, resembling unihemispheric sleep dynamics. Corresponding space-time plots of the system variable \(u_k\) confirm these findings.


The dynamics of the system are strongly influenced by two key frequency regions. One prominent synchronization region appears near the intrinsic frequency of the uncoupled FitzHugh-Nagumo oscillators. Remarkably, even though the external driving stimulus directly affects only two auditory nodes, the entire network tends to synchronize around this frequency for relatively small stimulation amplitudes. This points to a broad and smooth transition into global coherence driven by resonance with the system's natural dynamics.

In contrast, a second synchronization region emerges as a narrow tongue in the parameter space, starting at a slightly lower frequency. This region is marked by a sharp, abrupt onset of synchronization, resembling a first-order phase transition. Within this tongue, all nodes oscillate at the same average rate, although slight phase differences remain between them. These phase lags are small but significant, and they affect how the coupling between oscillators contributes to the overall network dynamics.

Theoretical analysis shows that when nodes share the same average frequency but are slightly out of phase, the cumulative effect of their phase differences can be described in terms of an effective time shift. This shift modulates the interaction terms between coupled oscillators and alters the timing of their oscillations. As a result, the collective period of the synchronized state becomes linearly dependent on this effective time shift.

Importantly, in incoherent states where phase differences are widely distributed, this effective time shift approaches zero, and the network's behavior is primarily governed by the natural frequency of the uncoupled oscillators. This duality explains why synchronization can occur at both the intrinsic system frequency and at a lower, amplitude-sensitive frequency range. Furthermore, increasing the stimulation amplitude amplifies the contribution of phase differences, effectively broadening the synchronization tongue linearly with amplitude.

\begin{figure}
\centering
\includegraphics[width=.85\linewidth]{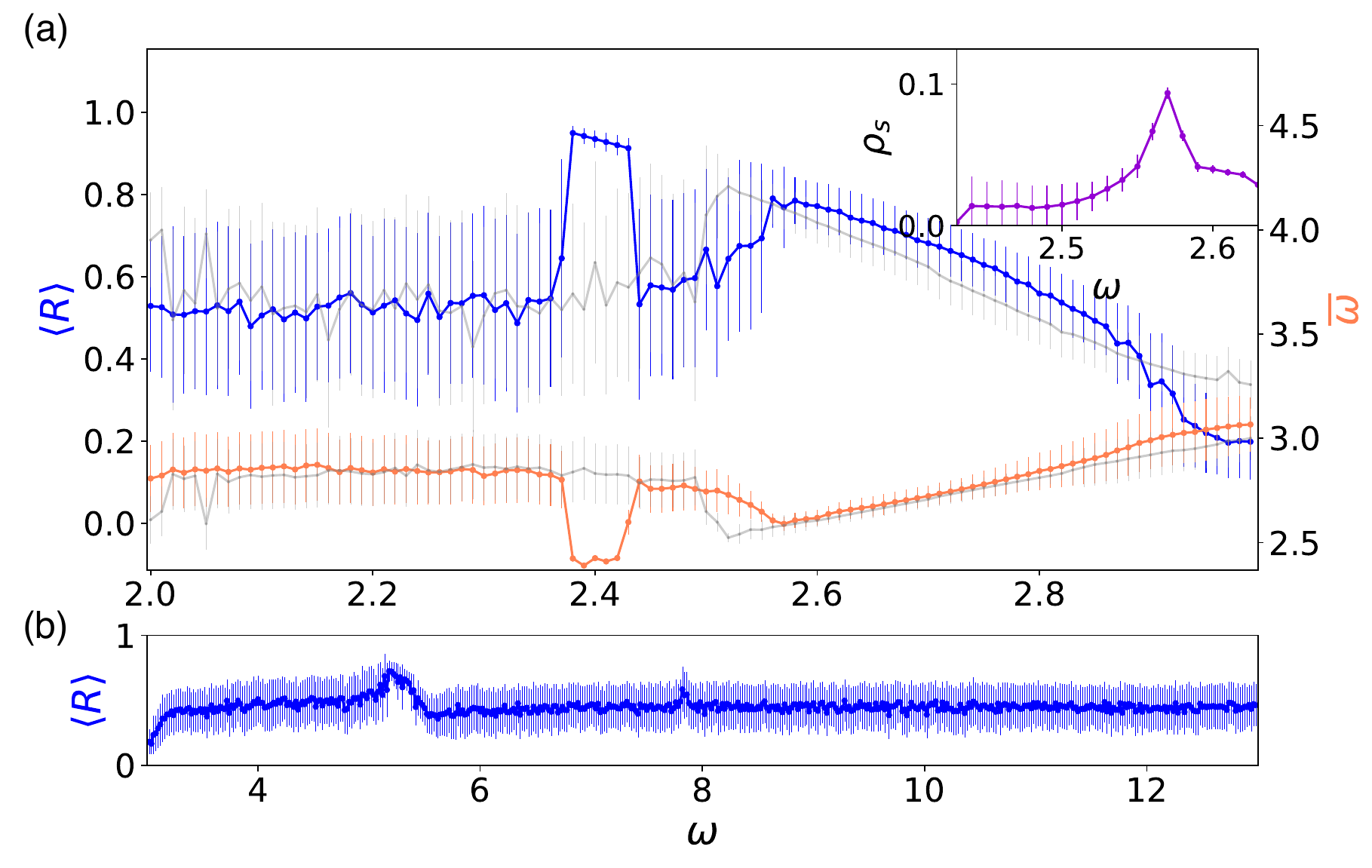}
\caption{Transition scenarios: (a) temporal mean of the Kuramoto order parameter $\langle R \rangle$ (dark blue) and the spatially averaged mean phase velocities $\overline{\omega}$ (light orange) in dependence on the frequency $\omega$ of the external stimulus for a fixed amplitude $\gamma=0.052$. The vertical bars indicate the standard deviation of the temporal mean of the Kuramoto order parameter and the spatially averaged mean phase velocities, respectively. As input nodes, the auditory cortices $k=41,86$ are chosen. In case of a different input ($k=1,45$) the corresponding light grey curves are shown in panel (a). The inset in panel (a) depicts $\rho_s=\frac{N_s}{\Delta T_L}$, the number $N_s$ of synchronized time intervals ($R(t)>0.8 \;\forall \,t$) divided by a simulation time of $\Delta T_L=30\,000$ for values of the frequency $\omega$ between the two synchronization regions. The vertical bars denote the standard deviation of the length of these synchronized time intervals. (b) $\langle R \rangle$ for a larger range of driving  frequencies $\omega$, showing higher resonance tonges. Other parameters are as in Fig.\,\ref{cha5:SAW21a:fig.2}. Figure taken from \cite{SAW21a}.}
\label{cha5:SAW21a:fig.4}
\end{figure}

In Fig.\,\ref{cha5:SAW21a:fig.4}a, both transitions are depicted in dependence on the frequency $\omega$ for a fixed amplitude $\gamma=0.052$. We can see an abrupt increase and decrease of the temporal mean of the Kuramoto order parameter $\langle R \rangle$ before and after $\omega\approx2.4$, respectively. In contrast, in approaching the upper synchronization region starting from $\omega\approx2.6$, $\langle R \rangle$ increases more slowly than at the transition to the synchronization tongue ($\omega\approx2.4$). In case of synchronization the standard deviation of $\langle R \rangle$, displayed by the vertical bars, is smaller than in case of desynchronized dynamics. That holds also for the spatially averaged mean phase velocities $\overline{\omega}$, which in case of synchronization takes over the lower value of the frequency $\omega$ of the external stimulus. Also for $\omega>2.6$, $\overline{\omega}$ is equal to $\omega$, whereas the standard deviation of $\overline{\omega}$ increases linearly with $\omega$. In contrast, there is no effect on the system for $\omega<2.4$. Neither $\langle R \rangle$ nor $\overline{\omega}$ show a different behavior for such values of $\omega$. The high value of the standard deviation of $\langle R \rangle$ stands for dynamics, where the Kuramoto order parameter $R(t)$ is fluctuating over its whole bandwidth $R \in [0,1]$. Simulations show that for $\omega>3.0$ the dynamical behavior of the system becomes similar to that with $\omega<2.3$. For both parameter intervals of $\omega$, there is no effect on the system. Simulations show also that a similar transition to synchronization at $\omega=2.6$ can be found for higher harmonics, i.e., multiple values of $\omega=2.6$. In Fig.\,\ref{cha5:SAW21a:fig.4}b, we can identify synchronization regions for $\omega=5.2, 7.8$ and $10.4$ becoming less pronounced for increasing $\omega$, i.e., having a smaller extension in the plane of $\omega$ and $\gamma$. In contrast, we could not detect repeated synchronization tongues of $\omega$ for multiple values of $\omega=2.4$. This indicates the existence of two different synchronization mechanisms.  

The existence of two synchronization regions depends on the choice to which nodes the external stimulus is supplied. In case of a different input, for instance $k=1,45$ in contrast to $k=41,86$, the light grey curves in Fig.\,\ref{cha5:SAW21a:fig.4}a depict the corresponding dependence of the Kuramoto order parameter $\langle R \rangle$ and the spatially averaged mean phase velocities $\overline{\omega}$ upon the frequency $\omega$ of the external stimulus. The synchronization region at $\omega\approx 2.4$ is missing here and only one synchronization region remains ($\omega > 2.5$).

The inset of Fig.\,\ref{cha5:SAW21a:fig.4}a confirms the increasing regularity between the two synchronization regions by depicting $\rho_s=\frac{N_s}{\Delta T_L}$ versus $\omega$, where $N_s$ is the number of synchronized time intervals ($R(t)>0.8 \, \forall \, t$) and $\Delta T_L=30\,000$ is the simulation time. The vertical bars denote the standard deviation of the length of these synchronized time intervals. With increasing $\omega$ not only the number of synchronized time intervals is increasing, but the standard deviation of their duration is decreasing. For $\omega>2.6$ we enter the synchronization region, where the value of $\rho_s$ drops due to the nearly consistently synchronized dynamics.

\section{Transitional role of auditory cortex}
\label{cha5:SAW24}

Synchronization patterns and coherence are widely recognized as fundamental to the operation of brain networks, playing pivotal roles in both healthy and pathological states. Notably, during auditory perception, there is an observable enhancement in coherence between the network's global dynamics and the auditory input. In this section, we demonstrate that synchronization phenomena are governed by a delicate interplay involving the network's topology, the spatial location of the input, and the frequency characteristics of the cortical input signals. To explore these dynamics, we investigate the effects of external stimulation on a network of FitzHugh-Nagumo oscillators configured with empirically derived structural connectivity. Furthermore, we examine the impact of cortical stimulation applied to various regions, with a particular focus on the auditory cortex \cite{SAW24}. 

In this section, the external input stimulus $I(t)=C_k^I\gamma \cos \omega t$ is modeled by a trigonometric function with frequency $\omega$ and amplitude $\gamma$ and is applied to a specific pair of cortical regions $k=I_0$ and $k=I_0+45$, where the index $I=I_0$ denotes the stimulated area. For instance, $k=41$ and $k=86$ are associated with the auditory cortex, i.e. $C_k^I=1$ if $k=41$ or $k=86$ and zero otherwise. The intra-hemispheric coupling between the single regions is given by the coupling strength $\sigma$, and the inter-hemispheric coupling is given by $\varsigma$. As we are looking for partial synchronization patterns we fix $\sigma = 0.7$ and $\varsigma=0.15$ similar to numerical studies of synchronization phenomena during unihemispheric sleep \cite{SCH21} and epileptic seizures \cite{GER20} where partial synchronization patterns have been observed. Given the uncertainty in the empirical connectivity data~\cite{SKO22}, note that the precise choice of the interhemispheric coupling is not crucial since the cortical input is chosen symmetrically in the corresponding areas of both hemispheres.

\begin{figure}
\centering
\includegraphics[width = .9\linewidth]{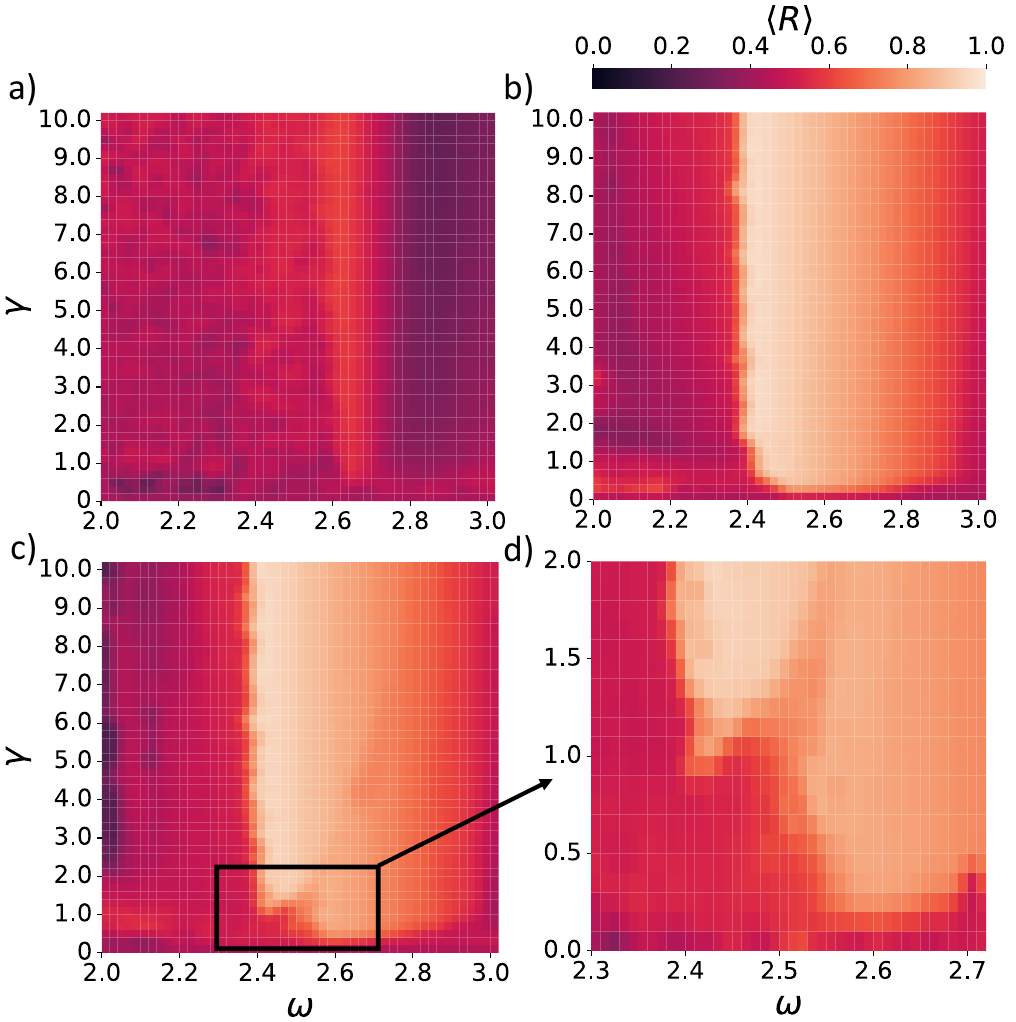}
    \caption{Synchronization regions in brain network with external stimulus: $\langle R \rangle$ in the parameter plane of the frequency $\omega$ and the input strength $\gamma$ of the external stimulus for cortical input regions (a) $k=14$, $I=1$, (b) $k=34$, $I=45$, and (c) $k=41$, $I=15$. Panel (d) shows a blow-up of (c). 
    Other parameters: $\sigma=0.7$, $\varsigma=0.15$, $\epsilon = 0.05$, $a=0.5$, and $\phi = \frac{\pi}{2} - 0.1$. Figure taken from \cite{SAW24}.}
    \label{cha5:SAW24:fig.2}
\end{figure}

We will now investigate the role of the auditory cortex in the collective dynamics of the human brain. For this purpose, we feed a periodic external input into specific areas of our neural network, using the regions in pairs as described in the AAL atlas (Table \ref{cha5:SAW24:tab:AAL}). Depending on the selected cortical regions $I$, a different influence on the degree of synchronization of the overall network can be observed as shown for three different input regions in Fig.\,\ref{cha5:SAW24:fig.2}. The light colored regions in Fig.\,\ref{cha5:SAW24:fig.2} indicate synchronized dynamics, whereas the darker colors indicate desynchronized dynamics. For $k=14$, $I=1$ in Fig.\,\ref{cha5:SAW24:fig.2}a, there is a slightly light colored stripe around the natural uncoupled frequency $\omega=2.6$ ($\langle R \rangle \approx 0.7$). For $k=34$, $I=45$ in Fig.\,\ref{cha5:SAW24:fig.2}b, there is a pronounced light colored synchronization region ($\langle R \rangle \approx 1$) starting at $\omega=2.4$. For $k=41$, $I=15$ in Fig.\,\ref{cha5:SAW24:fig.2}c, a triangular  synchronization tongue splits off from the bottom left of the broad synchronization region, starting at $\omega=2.4$. This is shown in more detail in the close-up in Fig.\,\ref{cha5:SAW24:fig.2}d. 

\begin{figure}
\centering
\includegraphics[width = .9\linewidth]{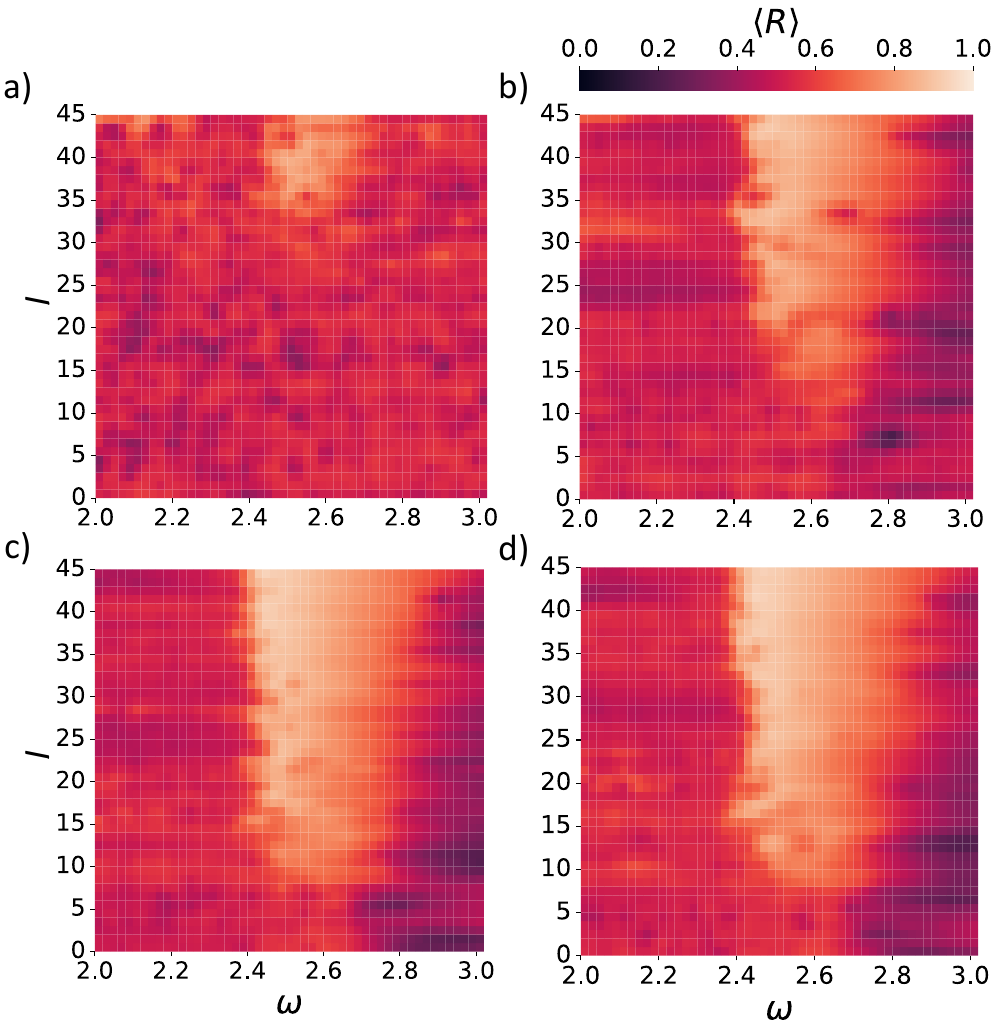}
    \caption{Same as in Fig.\,\ref{cha5:SAW24:fig.2}  in the parameter plane of the frequency $\omega$ of the external stimulus and its cortical input region $I$  (see Table \ref{cha5:SAW24:tab:AAL}) for input strengths (a) $\gamma=0.11$, (b) $\gamma=1.1$, (c) $\gamma=7.0$, and (d) $\gamma=11.0$. Other parameters as in Fig.\,\ref{cha5:SAW24:fig.2}. Figure taken from \cite{SAW24}.}
    \label{cha5:SAW24:fig.3}
\end{figure}

To further elaborate the different influence of different input regions, the global order parameter $\langle R \rangle$ is shown in Fig.\,\ref{cha5:SAW24:fig.3} in dependence of the frequency $\omega$ of the external stimulus and its cortical input region $I$ for four values of the input strength $\gamma$. In the case of the auditory cortex ($I=15$), a distinct influence on the neuronal network can be observed, which is not so pronounced for other regions. There is a threshold for global synchronization at input frequencies of $\omega = 2.4$. Even for very small input strengths $\gamma=0.11$ (see Fig.\,\ref{cha5:SAW24:fig.3}a), synchronization of the entire network can be achieved for certain input regions $I$. On the other hand, even for very large input strengths $\gamma=7.0$ (see Fig.\,\ref{cha5:SAW24:fig.3}c), some input regions never induce synchronization of the entire system. For better visibility, the input region indices $I ={\cal P}(1,...,N)$ (modulo 45) are permuted according to their synchronizability by re-arranging them in ascending order of the row sum of the temporal mean of the Kuramoto order parameters $\langle R \rangle$ averaged over all 4 panels of Fig.\,\ref{cha5:SAW24:fig.3}. Increasing the input strength further to $\gamma=11.0$ (see Fig.\,\ref{cha5:SAW24:fig.3}d) does not lead to any quantitative change of the synchronization tongue.

Interestingly, the auditory cortex $I = 15$ (relabeled, marked yellow in Table \ref{cha5:SAW24:tab:AAL}) is an input region that allows for the synchronization of the entire system at sufficiently large $\gamma$, but not at very small $\gamma$ (Fig.\,\ref{cha5:SAW24:fig.3}a). In this sense it plays an intermediate role between cortical areas which can easily synchronize the brain (marked pink in Table \ref{cha5:SAW24:tab:AAL}) and those which never synchronize (marked blue in Table \ref{cha5:SAW24:tab:AAL}). 

\begin{table}[tbp!]
\centering
\begin{tabular}{|c|c|c|c|} 
   \hline
   $k$: L/R & $I$ & Region &Lobe \\
   \hline
   \hline
   1/46  & 35 & Precentral &  Central region\\
   2/47  & 18 &Frontal Sup &  Frontal lobe\\
   \rowcolor{cyan}3/48  & 7 & Frontal Sup Orb & Frontal lobe\\
   4/49  & 31 &Frontal Mid &  Frontal lobe\\
   \rowcolor{cyan}5/50  & 8 & Frontal Mid Orb & Frontal lobe\\
   6/51  & 34 &Frontal Inf Oper &Frontal lobe \\
   7/52  & 29 &Frontal Inf Tri  &  Frontal lobe\\
   \rowcolor{cyan}8/53  & 3 &Frontal Inf Orb &  Frontal lobe\\
   9/54  & 11 &Rolandic Oper &  Central Region \\
   10/55 & 30 &Supp Motor Area &  Frontal lobe\\
   \rowcolor{cyan}11/56 & 5 &Olfactory & Frontal lobe\\
   12/57 & 33 &Frontal Sup Medial &  Frontal lobe\\
   13/58 & 21 &Frontal Med Orb  &  Frontal lobe\\
   \rowcolor{cyan}14/59 & 1 & Rectus & Frontal lobe\\
   15/60 & 22 &Insula  & Insula\\
   16/61 & 20 &Cingulum Ant  & Limbic lobe\\
   17/62 & 25 &Cingulum Mid & Limbic lobe\\
   18/63 & 26 &Cingulum Post & Limbic lobe\\
   19/64 & 13 &Hippocampus  & Limbic lobe \\
   20/65 & 16 &ParaHippocampal & Limbic lobe\\
   \rowcolor{cyan}21/66 & 2 &Amygdala & Sub cort gray nuc\\
   \rowcolor{pink}22/67 & 39 &Calcarine  & Occipital lobe\\
   \rowcolor{pink}23/68 & 43 &Cuneus  & Occipital lobe\\
   \rowcolor{pink}24/69 & 37 &Lingual  & Occipital lobe\\
   \rowcolor{pink}25/70 & 42 &Occipital Sup & Occipital lobe \\
   \rowcolor{pink}26/71 & 38 &Occipital Mid  & Occipital lobe\\
   27/72 & 32 &Occipital Inf  & Occipital lobe \\
   28/73 & 23 &Fusiform   & Occipital lobe\\
   \rowcolor{pink}29/74 & 41 &Postcentral & Central region\\
   30/75 & 28 & Parietal Sup & Parietal lobe \\
   31/76 & 36 & Parietal Inf  & Parietal lobe\\
   \rowcolor{pink}32/77 & 44 &Supramarginal & Parietal lobe \\
   \rowcolor{pink}33/78 & 40 &Angular  & Parietal lobe\\
   \rowcolor{pink}34/79 & 45 &Precuneus & Parietal lobe\\
   35/80 & 14 &Paracentral Lobule & Frontal lobe\\
   \rowcolor{cyan}36/81 & 9 &Caudate & Sub cort gray nuc\\
   37/82 & 19 &Putamen &Sub cort gray nuc\\
   38/83 & 12 &Pallidum &  Sub cort gray nuc\\
   39/84 & 17 &Thalamus & Sub cort gray nuc\\
   40/85 & 10 &Heschl &  Temporal lobe\\
   \rowcolor{yellow}41/86 & 15 &Temporal Sup & Temporal lobe\\
   \rowcolor{cyan}42/87 & 4 &Temporal Pole Sup & Limbic lobe\\
   43/88 & 24 &Temporal Mid & Temporal lobe\\
   \rowcolor{cyan}44/89 & 6 &Temporal Pole Mid & Limbic lobe\\
   45/90 & 27 &Temporal Inf &Temporal lobe \\ \hline
\end{tabular}

\caption{Cortical and subcortical regions \cite{TZO02} as introduced in Tab.\,\ref{cha5:SAW24:tab:AAL}. $I$ denotes the relabeled index of the regions ordered according to increasing influence on global synchronization. A small input signal ($\gamma = 0.11$, see Fig.\,\ref{cha5:SAW24:fig.3}a) to the pink shaded regions has a big impact on the synchronization of the whole system ($\langle R \rangle > 0.8)$, whereas even strong inputs ($\gamma = 11.0$, see Fig.\,\ref{cha5:SAW24:fig.3}d) to the blue shaded regions have no impact on synchronization ($\langle R \rangle < 0.5)$. The brain areas $k=41,86$ (yellow shaded, $I=15$) are associated with the auditory cortex.   
}
\label{cha5:SAW24:tab:AAL}
\end{table}

The system's response to external stimulation varies significantly depending on the cortical region targeted. With constant amplitude and frequency beyond the synchronization threshold, three distinct dynamical regimes emerge. When input is applied to regions that are less influential in network-wide coordination, the system exhibits desynchronized dynamics similar to its behavior without external driving. The Kuramoto order parameter fluctuates strongly over time, and only local pockets -- primarily in the right hemisphere -- show limited synchrony, while most nodes, especially in the left hemisphere, remain asynchronous. In contrast, stimulation of the auditory cortex leads to an intermediate state characterized by alternating periods of synchronization and desynchronization. During synchronized episodes, the mean phase velocities of the nodes align closely with the collective frequency, although this coherence is not sustained. When input is delivered to regions more effective in orchestrating global dynamics, the system exhibits strong and persistent synchronization, with phase velocities across all nodes matching the frequency of the external stimulus.

These differences are also reflected in the spatial phase dynamics. Input to less effective regions results in spatially irregular patterns, whereas stimulation of the auditory cortex generates alternating coherent and incoherent patterns. In the fully synchronized regime, phase alignment is both spatially uniform and temporally stable. The auditory cortex, therefore, plays a transitional role, capable of promoting but not fully maintaining global synchrony. This sensitivity to input and the coexistence of order and disorder suggest that the system operates near a critical regime, where small perturbations can tip the dynamics between synchronization and desynchronization.

\section{Neural network model and music}%
\label{cha5:SAW22}

This section presents a computational study using FitzHugh-Nagumo oscillators on empirically derived brain networks to investigate how music influences brain dynamics. The results show that musical input enhances neural coherence, especially in the gamma-band range, which is linked to cognitive processing. High-frequency components, particularly in the gamma range, most effectively drive synchronization, consistent with empirical findings of increased gamma coherence during music listening at cognitively salient moments~\cite{SAW22}.

A functional separation is observed: high frequencies are associated with cortical cognitive activity, while low frequencies relate to subcortical coordination and motor responses, supporting theories of distinct neural systems for different aspects of music perception. Low-frequency synchronization, in particular, reflects rhythmic entrainment and links to motor activity, highlighting the integrative role of music in engaging both sensory and motor circuits. These findings align with EEG and fMRI evidence and offer a computational framework for understanding how auditory stimuli shape large-scale brain dynamics. This perspective article underscores music's potential to modulate neural coherence, with implications for therapeutic applications in neurology~\cite{SAW22}.

Sound is transformed into neural spikes through mechanical-to-electrical transduction in the cochlea, which acts as a frequency analyzer due to spatial variations in stiffness along the basilar membrane. The transformation of sound into neural spikes is the subject of much current research \cite{TRI10, SCH11k, MIZ14, BAD15, BAD17, BAD18, GUO21, SAW22}. Traveling waves induced by sound pressure cause localized membrane displacements, which, when meeting specific spatial and temporal criteria, trigger neural spikes via stereocilia. These spikes are organized into 24 critical bands and relayed through the auditory pathway, preserving tonotopy up to the cortex. A biophysically detailed model using a rod-based Finite-Difference Time Domain approach simulates this process by solving a position-dependent differential equation for membrane motion, driven by digital sound input. The model generates a time series of spike outputs, replicating observed auditory processing features such as frequency selectivity, phase synchronization, and delay patterns based on frequency, providing a mechanistic basis for studying auditory perception.

\begin{figure}[tbp!]
    \begin{center}
\includegraphics[width = .8\textwidth]{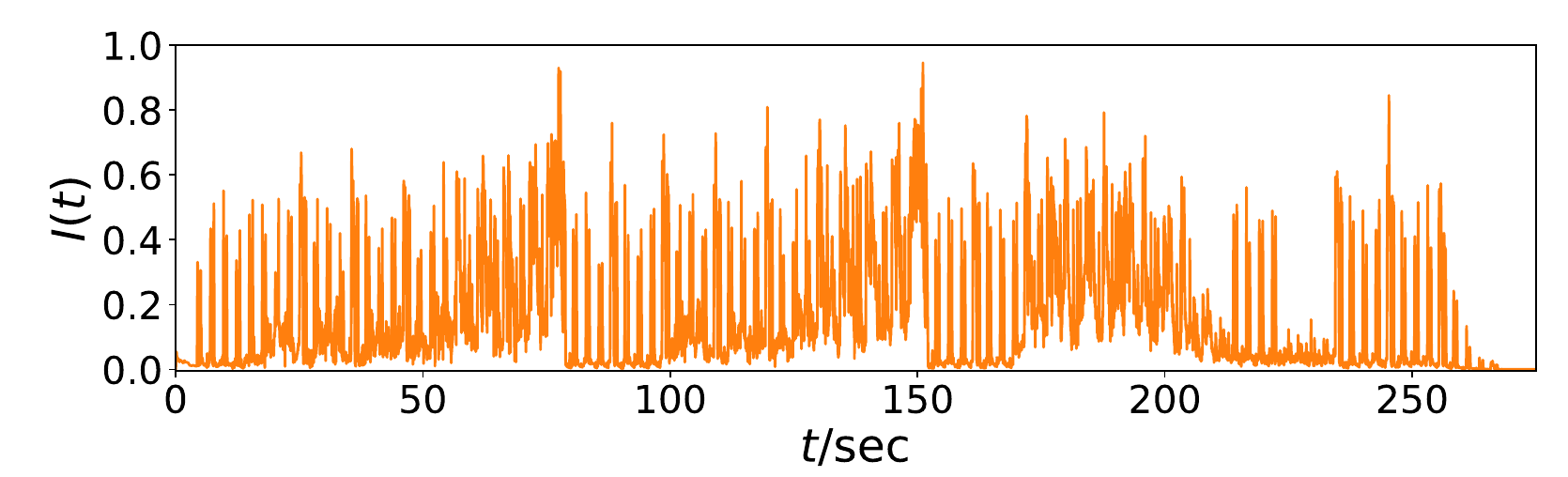}
    \caption{Time-series of the neural input signal $I(t)$ obtained from the music song \textit{One Mic} transformed by a method developed by Bader \cite{BAD20}. The song has a length of about 270 seconds and was released in 2002 by American rapper \textit{Nas}. Figure taken from \cite{SAW22}.}
    \label{cha5:SAW22:fig.2}
    \end{center}
\end{figure} 

The auditory cortex, located bilaterally in the temporal lobes, plays a central role in processing auditory information, involving both basic and complex hearing functions. It participates in spectrotemporal analysis of auditory input received from the ear. Figure~\ref{cha5:SAW22:fig.2} illustrates the time-series of neural impulses generated by the auditory cortex in response to auditory stimuli. These impulses were obtained using Bader's method, as described above \cite{BAD15, BAD17, BAD18}, and were recorded at a sampling rate of 192 kHz. For the simulations, a real music piece, ``One Mic'' by Nas, was used as the auditory stimulus.

The external stimulus, \(I(t)\), corresponds to the neural impulses induced by the music and is applied to the brain regions associated with the auditory cortex (locations \(k=41,86\) in the AAL index). The FitzHugh-Nagumo model operates in dimensionless time, so to align the model's time units with real time, the oscillation period of a single FHN oscillator is compared with the characteristic frequencies of the empirical time series. The real time units are then derived by scaling the simulation time using the formula \( f_b = n_b / f_{\text{FHN}} \), where \(n_b\) represents the frequency in Hz, and \(f_{\text{FHN}} \approx 0.4\) is the dimensionless frequency of the FHN oscillator \cite{BAD15, BAD17, BAD18}.

Additionally, we introduce a new measure which specifies the coherence between the Kuramoto order parameter and the input signal by using the time average of the Kuramoto order parameter weighted with the input signal

\begin{equation}
\gamma = \frac{1}{\Delta T}\int_0^{\Delta T} R(t)I(t) \,\mathrm{d} t
\label{cha5:SAW22:eq:gamma}
\end{equation}

to quantify the overlap of coherent episodes ($R$ large) with large input signals, averaged over time. The coherence $\gamma$ is maximum if the synchronization is large whenever the signal is large. It is small if the overall synchronization is low, or if the modulation of the synchronization in time is not in phase with the modulation of the input signal amplitude. For $\gamma=0$ the Kuramoto order parameter and the input signal do not overlap at any time point. An increased value of $\gamma \in [0,1]$ means increased overlap  between the Kuramoto order parameter and the input signal. The motivation for introducing the measure $\gamma$ lies in the fact that in the human brain the increase and decrease of synchronization follows the large-scale form of the listened music in a coherent way \cite{HAR14, HAR20a}.

It has been examined how the frequency of external stimuli, specifically music, influences coherence in the auditory cortex \cite{SAW22}. This perspective article finds that coherence peaks in the gamma-band range (30-120 Hz), aligning with previous research on music perception. As the frequency of the stimulus increases, the brain's synchronization with the stimulus decreases. At 30 Hz, the auditory cortex's dynamics most closely match the stimulus, with temporal modulations in the system reflecting those of the music. Additionally, hemispheric differences in phase velocity are observed, with the right hemisphere synchronizing while the left remains desynchronized, similar to patterns seen in unihemispheric sleep.

\section{Comparison with experiments}
\label{cha5:SAW22:6}

The coupling of oscillatory neural signals within traditional frequency bands is considered a key mechanism underlying various perceptual, sensorimotor, cognitive, and motor functions. These include Gestalt perception and binding \cite{GRA89c, TAL95, KEI99, TAL99, ROD99a, ENG01b, ENG01c}, timing and expectation \cite{BUH05, BUH09}, attention \cite{WOM07a, FRI09c, NIK13a}, consciousness \cite{BAA06, DEH11b, ENG16, OWE19}, motor functions \cite{THA15}, and music perception \cite{BHA01, ZAN05}. According to \cite{ENG16}, brain activity typically clusters into frequency bands: delta (0.5–3.5 Hz), theta (4–7 Hz), alpha (8–12 Hz), beta (13–30 Hz), and gamma (>30 Hz). Gamma-band activity, being the newest area of interest since the late 1990s, has varying definitions across studies. For consistency, we adopt the classification from \cite{FRE13a}, which distinguishes between low gamma (30–60 Hz) and high gamma (60–120 Hz). Frequencies above 120 Hz are referred to as 'fast oscillations' following \cite{BUZ06}. The gamma-band is especially significant for large-scale synchronization, as it is believed to play a crucial role in integrating information across different cortical regions.

The perception of musical form, as the highest hierarchical level of musical structure, is influenced by various cognitive processes, including Gestalt laws, which group notes, bars, and phrases into higher-level structures such as verses, choruses, and classical sonata forms \cite{LER90, HAR20a, LEM97a, DEU13, NEU13, DEL14}. This hierarchical structuring creates contrasts within the musical form, such as the tension and relaxation experienced in different musical genres (e.g., Techno, Classical) \cite{KOE14, LEH15c}. The transition from expectation (potential energy) to action (kinetic energy) in music, such as dancing, aligns with neural entrainment, where auditory cortex neurons synchronize with motor cortex neurons \cite{KUR31, THA15}. Computational methods in music information retrieval, such as analyzing amplitude, spectral centroid, and fractal correlation dimension, help reveal these contrasts in musical structure. These properties are linked to the perceived loudness, brightness, and complexity of music, which influence the dynamics of musical form \cite{GRA83, GRA83a, BAD13, HAR20a, BAD21, LIN21b, BAD21a}. Based on these observations, this perspective article hypothesizes that neural synchronization will correspond to the amplitude of the musical stimulus, with higher synchronization during high amplitude (louder) sections of music, particularly in the gamma-band, which plays a significant role in the perception of musical parameters.

In an experiment examining the perception of large-scale musical form, electroencephalogram (EEG) recordings were obtained from human participants, adhering to ethical guidelines including informed consent \cite{SAW22}. Participants were recruited primarily through the Institute of Systematic Musicology Hamburg, and had instrumental training or DJ experience (mean duration: 10.0 years, standard deviation: 4.6 years). A total of 25 musically skilled subjects listened to the song ``One Mic'' by Nas, which was released in 2001. The EEG signals were recorded at a sampling rate of 500 Hz from 32 electrodes positioned according to the 10-20 system \cite{JAS58}. Following artifact correction, the data for each channel were averaged across subjects and trials to create a grand average of 75 trials per channel, which increased the signal-to-noise ratio and enhanced event-related potentials (ERPs). This averaging approach allowed for the analysis of evoked potentials, as opposed to induced potentials, in response to the stimulus, in line with classical ERP methodologies \cite{TAL96, TAL99, ZAN05}. The consistency in electrode positions throughout the measurements ensured that the observed differences in correlation strength between electrodes were not due to spurious synchrony \cite{HOL77, KAY06, BHA18}. Further details regarding the experimental procedure, technical aspects, and preprocessing can be found in \cite{HAR20a}.

The EEG signals from all channels were decomposed into nine independent frequency bands using a continuous wavelet transform with a Mexican Hat wavelet \cite{FRE13a}. This method is efficient compared to using a bandpass filter followed by a Hilbert transform, as it allows for direct decomposition of the EEG data into the desired frequency bands by defining the number of octaves. The wavelet transform provides a time-frequency representation where each frequency band corresponds to a specific musical octave, and the scales are adjusted in a way similar to an equal-tempered musical scale, where each octave doubles the frequency of its lower counterpart. This alignment between EEG bands and musical scales, while potentially coincidental, could also reflect the logarithmic nature of human sensory perception \cite{SCH18l}.

\begin{figure}
\centering
\includegraphics[width = 1.0\textwidth]{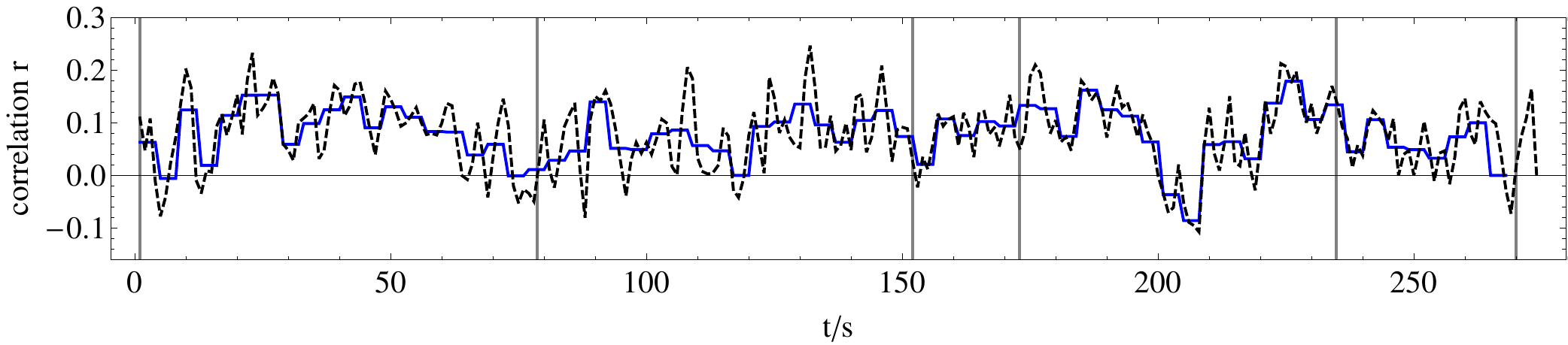}
    \caption{Example of the synchronization dynamics between two electrodes. Dashed black line: Time series of the Pearson correlation coefficient $r$ calculated for successive 1-second time windows ($n = 500$ between averaged EEG recordings of electrode Fp1 and electrode T7. Blue line: Pearson correlation coefficient averaged over 4 consecutive 1-second time windows of the dashed black line. Figure taken from \cite{SAW22}.
		}
    \label{cha5:SAW22:fig.7}
\end{figure}

Figure \ref{cha5:SAW22:fig.7} illustrates the synchronization dynamics between two electrodes. The plot shows the Pearson correlation coefficient calculated for successive 1-second time windows between EEG signals from electrodes Fp1 and T7. The dashed black line represents the time series of the correlation coefficient, while the blue line depicts the moving average of this coefficient over four consecutive 1-second time windows. This approach provides insights into the temporal synchronization between different brain regions during the music perception task.

For each electrode pair across the nine frequency bands, synchronization dynamics were analyzed by calculating the linear cross-correlation, using the Pearson correlation coefficient \(r\), which measures the strength of the correlation between two time series or variables \cite{GLA02,BAS15c,ERR17}. This method is widely employed for its simplicity and efficiency in quantifying synchronization without the need for time delays. The Pearson correlation is calculated within successive 1-second time windows for each possible electrode pair within each wavelet-filtered dataset, resulting in 4,464 time series, each of 270 seconds corresponding to the stimulus duration (see Figure \ref{cha5:SAW22:fig.7}).

This approach provides a detailed view of synchronization dynamics across different frequency bands in the context of musical form perception. The Pearson correlation coefficient offers a practical and effective means to assess linear correlations, yielding results consistent with other synchronization measures such as the phase-locking value \cite{LAC99}. Further comparisons of synchronization techniques are discussed in \cite{JAL14b}.

\begin{figure}
\centering
\includegraphics[width = .9\textwidth]{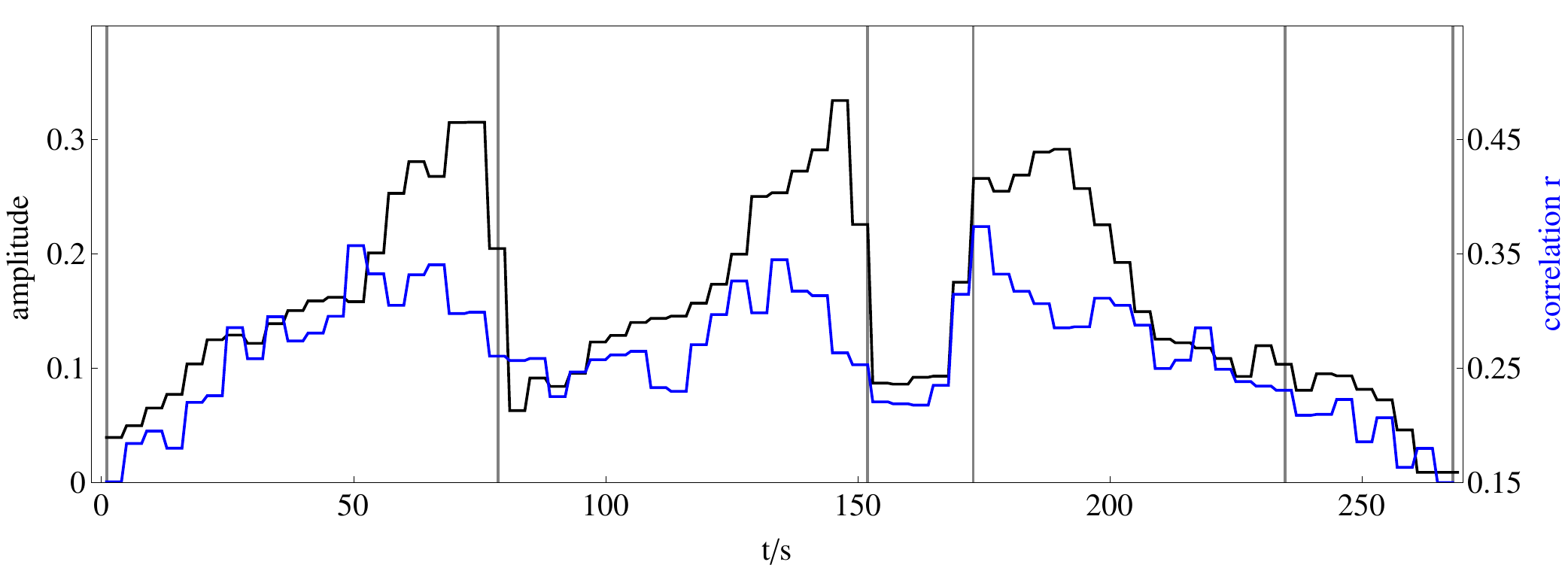}
    \caption{Comparison of whole brain synchronization dynamics and representation of the musical form of the stimulus. The black line shows the amplitude dynamics of the stimulus as a representation of the musical form, averaged over each of 4 consecutive seconds. The blue line shows the average of the 25 correlation time series between two electrodes from each frequency band that correlates most strongly with the amplitude dynamics of the stimulus. Figure taken from \cite{SAW22}.}
    \label{cha5:SAW22:fig.8}
\end{figure}

To relate the synchronization dynamics of the EEG time series to the amplitude modulation of the stimulus, the amplitude of the stimulus and the correlation coefficients for the 496 electrode pairs across the nine frequency bands were averaged within successive 4-second windows. This step helps avoid minor fluctuations and aligns with changes related to the musical form. The correlation between these averaged time series and the amplitude time series of the stimulus was then calculated. The 25 time series per frequency band that showed the strongest correlation with the stimulus amplitude were selected, averaged, and correlated with the amplitude dynamics of the stimulus. The results showed strong correlations between the low and high gamma bands (FB 2 and FB 3) as expected, as well as significant correlations in the slower oscillations (FB 7–9). These findings suggest that synchronization dynamics in each frequency band are strongly related to the amplitude dynamics of the stimulus.

Further analysis revealed that when the time series from all frequency bands were averaged, the correlation with the stimulus was maximized, with a Pearson coefficient of 0.76. This suggests that increased stimulus amplitude correlates with higher synchronization across the most correlated time series of different frequency bands, indicating a nonlinear network process in the brain related to sound perception. The correlation between synchronization and stimulus amplitude is not trivial, as brain synchronization occurs at frequencies much lower than the musical frequencies and across multiple perceptual parameters. Increases in sound amplitude lead to enhanced synchronization, emphasizing the nonlinear nature of this process.

The correlation patterns showed two distinct regimes, with high correlations in the frequency bands FB 2 and FB 3, and a frequency band (FB 5) with low correlation. The dynamics in FB 6–9, related to walking and dancing, are attributed to the interaction between the neocortex and subcortical brain regions. High correlations in the gamma bands (FB 2–3) suggest neocortical activity, while the high correlations in bands FB 6–9 reflect subcortical involvement. These results point to a separation in synchronization related to musical form, with cortical regions (FB 2–3) responsible for one part of the synchronization, and subcortical regions (FB 6–9) responsible for another.

Further analysis of the gamma-band (31.25–125 Hz) showed that these bands correspond to a frequency range with the strongest coherence between neural synchronization and external input. Both experimental and simulated results demonstrated a pronounced maximum correlation in the gamma-band (FB 2 and FB 3), with the second maximum observed in the experimental data due to the interaction of the neocortex with subcortical regions, which was absent in the simulated data. This suggests that subcortical brain activity contributes significantly to the synchronization dynamics observed in the experimental setup.

\begin{figure}
\centering
\includegraphics[width = .9\textwidth]{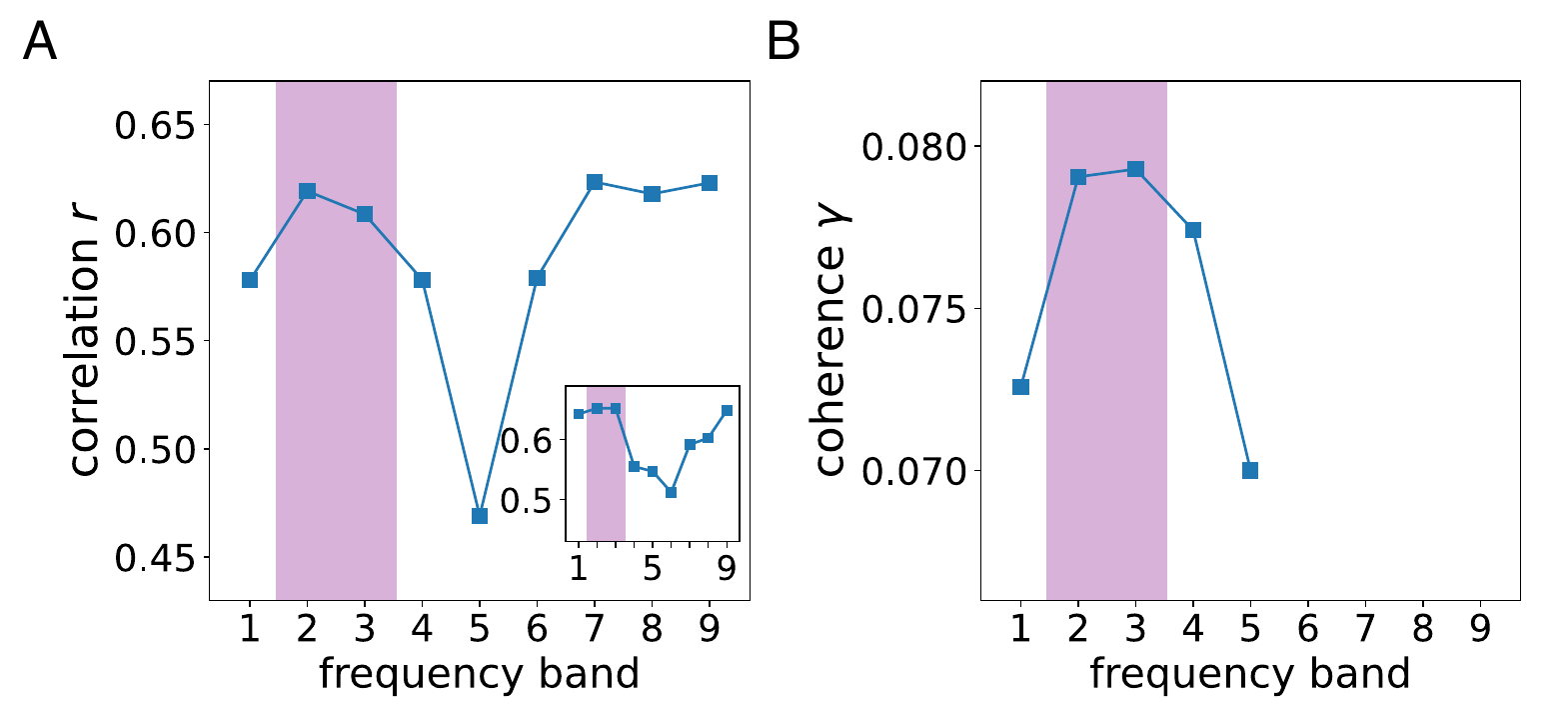}
    \caption{Comparison between experimental and numerical results: (A) Experimentally recorded correlation $r$ of the individual averages of the amplitude time series for each frequency band most strongly correlated with the stimulus as a function of frequency band (FB) FB 1: $125 - 250$ Hz, FB 2: $62.5 - 125$ Hz, FB 3: $31.25 - 62.5$ Hz, FB 4: $15.63 - 31.25$ Hz, FB 5: $7.81 - 15.63$ Hz, FB 6: $3.91 - 7.81$ Hz, FB 7: $1.95 - 3.91$ Hz, FB 8: $0.98 - 1.95$ Hz, FB 9: $0.49 - 0.98$ Hz. The inset depicts the Pearson correlation coefficient $r$ as a function of frequency band where instead of the amplitude the fractal dimension \cite{GRA83,GRA83a} has been used for the calculation of $r$. (B) Numerically simulated coherence $\gamma$ between network dynamics and external stimulus. The purple shaded regions in both panels indicate the gamma-band ($f_b \approx 30 - 120$ Hz), respectively. Figure taken from \cite{SAW22}.}
    \label{cha5:SAW22:fig.9}
\end{figure}

\section{Conclusion and outlook}%
\label{cha5:SAW22:7}

This perspective article explores the impact of auditory stimuli on neural network dynamics using the FitzHugh-Nagumo (FHN) model, integrated with empirical human brain connectivity data derived from diffusion-weighted MRI. The primary objective is to understand how synchronization in brain networks is influenced by external auditory input, with a focus on the frequency and amplitude of the stimuli. The results demonstrate that synchronization in the network is highly sensitive to the frequency of the input, particularly when the input frequency aligns with the system's natural frequencies or the collective frequency of the coupled network. Synchronization levels increase significantly as the input frequency approaches these intrinsic frequencies or their harmonics.

In addition to frequency sensitivity, this perspective article finds that the amplitude of the auditory stimulus, representing its perceived volume, modulates synchronization. An increase in amplitude broadens the frequency range over which synchronization can occur, thereby stabilizing the synchronous states within the network. This finding suggests that auditory input not only triggers synchronization but also influences the extent and stability of the synchronized state, which is crucial for maintaining coherent brain activity during complex cognitive tasks.

This perspective article further investigates transition dynamics, revealing that the precise tuning of the input frequency can induce significant shifts in network synchrony, supporting the idea that synchronization in the brain is dynamic and context-dependent. These findings align with previous empirical studies, including those by Bader's group, which demonstrate that music induces alternating periods of neural synchrony and desynchrony. These transitions are interpreted as manifestations of the brain operating near criticality, where the system resides at the boundary between order and disorder. Such critical dynamics are associated with complex phenomena, including hysteresis and neuronal avalanches, which have been linked to phase transitions in neural systems.

The results suggest that the brain's ability to alternate between synchronized and desynchronized states in response to auditory input may be essential for higher cognitive functions such as attention, temporal prediction, and memory processing. This perspective article highlights the importance of dynamic flexibility in brain networks, especially in the context of complex auditory stimuli like music. This flexibility likely facilitates the brain's ability to process and respond to changing external environments, making it adaptable to different forms of auditory input.

Moreover, this perspective article provides a theoretical foundation for understanding the neural impact of music on brain networks, demonstrating how simplified auditory stimuli can induce complex synchronization patterns in large-scale brain networks. This has significant implications for understanding the neural mechanisms underlying music-related cognitive processes, such as memory formation, attentional modulation, and emotional engagement, as well as for investigating how music influences brain states in both healthy and clinical populations.

This perspective article also uses a biologically informed cochlear model, which simulates the frequency-specific transduction of acoustic input into spike-like signals via the basilar membrane. These signals were used to drive the FHN oscillator network, which reflects the structural organization of the brain as mapped by diffusion MRI. The simulation results reveal a peak in synchronization, particularly in the gamma-band, and show that this synchronization corresponds to transitions in the structure of the musical input. Specifically, synchronization is most pronounced at the transitions between large-scale musical segments, which are termed musical high-level events. This pattern suggests that the brain's neural coherence in the gamma-band is closely tied to the perception and cognitive structuring of music.

The results of the simulation are consistent with empirical EEG data, which also show that music listening induces alternating states of synchrony and desynchrony in brain activity. The analysis demonstrates that these fluctuations in synchrony are closely linked to the temporal structure of the music, reinforcing the idea that musical form has a direct impact on brain network dynamics. Moreover, statistical analyses such as Pearson correlation of the sound envelope and fractal correlation dimension confirmed that the strongest correspondence between network dynamics and auditory input occurs within the gamma-band. These findings align with theoretical models of music perception, which suggest that high-frequency oscillations play a crucial role in the brain's ability to process and organize musical structure.

In addition, this perspective article reveals a functional dissociation between frequency bands, with high-frequency synchronization predominantly associated with neocortical processing, while low-frequency activity appears to reflect interactions between cortical and subcortical regions. This distinction is particularly evident in the low-frequency bands, which are linked to sensorimotor functions related to rhythm and movement. This functional division emphasizes the role of different brain regions in processing various aspects of auditory input, from basic sensory perception to more complex cognitive and motor responses.

The observed alternation between synchronized and desynchronized states suggests that the brain operates near a critical state, a phenomenon marked by dynamic variability and complex transitions between ordered and disordered states. Such criticality is widely recognized in neural systems and is thought to optimize information processing, as it enables the system to be highly sensitive to small changes in input. This study supports the hypothesis that the brain functions near a critical point, where synchronization dynamics can adaptively respond to changing stimuli.

Future research directions may include incorporating more realistic musical stimuli and exploring the effects of different acoustic features (such as tempo, harmony, and rhythm) on brain synchronization. Expanding the model to include more detailed brain parcellations and integrating other sensory modalities could provide further insights into the neural mechanisms underlying music perception and multisensory integration. Additionally, applying this framework to clinical populations could offer valuable perspectives on the therapeutic potential of music-induced synchronization in conditions such as dementia, Parkinson's disease, and other neurocognitive disorders.

Overall, this study offers a computational framework for understanding how auditory input, brain connectivity, and critical dynamics interact to shape neural synchronization across brain networks. These findings contribute to a deeper understanding of the neural underpinnings of music perception and cognitive processing, providing a foundation for future investigations into the complex relationship between music and the brain.

\section{Acknowledgments}
The author acknowledges Rolf Bader, Lenz Hartmann, and Eckehard Sch{\"o}ll for stimulating discussion and collaboration. Special thanks are extended to the patrons of the ``Creative Performance Workshop'', whose support and platform enabled the presentation and synthesis of the results discussed herein, thereby facilitating the articulation of the broader perspectives outlined in this article.

\bibliography{ref}
\bibliographystyle{tfs}

\end{document}